\documentclass[11pt]{article}

\usepackage[preprint]{acl}

\usepackage{times}
\usepackage{latexsym}

\usepackage[T1]{fontenc}
\usepackage[utf8]{inputenc}

\usepackage{microtype}

\usepackage{inconsolata}

\usepackage{graphicx}
\usepackage{booktabs}
\usepackage{multirow}
\usepackage{subcaption}
\usepackage{amssymb}
\usepackage{xcolor}
\usepackage{pifont}

\title{OpenSTBench: Beyond Semantic Evaluation for Speech Translation}

\author{
 \textbf{Yanjie An\textsuperscript{1}\thanks{These authors contributed equally to this work.}},
 \textbf{Yuxiang Zhao\textsuperscript{1}\footnotemark[1]},
 \textbf{Yichi Zhang\textsuperscript{1}},
 \textbf{Qixi Zheng\textsuperscript{1}},
\\
 \textbf{Yujie Tu\textsuperscript{2,4}},
 \textbf{Keqi Deng\textsuperscript{3}},
 \textbf{Kai Yu\textsuperscript{1}},
 \textbf{Xie Chen\textsuperscript{1,2}\thanks{Corresponding author.}}
\\
 \textsuperscript{1}MoE Key Lab of Artificial Intelligence, Jiangsu Key Lab of Language Computing,\\
X-LANCE Lab, School of Computer Science, Shanghai Jiao Tong University, Shanghai, China,\\
 \textsuperscript{2}Shanghai Innovation Institute, Shanghai, China,
 \textsuperscript{3}Microsoft, USA,\\
 \textsuperscript{4}University of the Chinese Academy of
Sciences, Beijing, China
}

\begin{document}
\maketitle

\begin{abstract}

Speech translation systems increasingly span speech-to-text translation (S2TT), speech-to-speech translation (S2ST), offline translation, and streaming generation, producing outputs that differ in modality, speech realization, and timing behavior. Existing evaluation practices assess important aspects such as translation quality, speech quality, and temporal quality, but these aspects are often evaluated under separate protocols, making it difficult to compare heterogeneous systems comprehensively. To address this gap, we present OpenSTBench, a unified multidimensional evaluation framework that organizes heterogeneous speech translation outputs into a shared evaluation format. OpenSTBench supports both S2TT and S2ST systems in offline and streaming settings, and jointly evaluates translation quality, speech quality, speaker preservation, emotion and paralinguistic fidelity, temporal consistency, and latency. Through experiments on representative speech translation systems, we show that systems with strong translation quality can still differ substantially in speech quality, as well as in temporal quality. OpenSTBench provides a reproducible protocol for analyzing these cross-dimensional differences and supporting application-oriented comparison of speech translation systems. The code and datasets are available at \url{https://github.com/sjtuayj/OpenSTBench}.

\end{abstract}
\section{Introduction}

Speech translation is expanding from offline speech-to-text translation to settings that include speech-to-speech generation, spoken interaction, multilingual communication, and real-time translation. In these settings, system behavior is not fully described by translated text alone: target speech may differ in naturalness, speaker characteristics, expressive information, acoustic events, duration structure, and response latency.

Translation metrics such as BLEU~\cite{papineni-etal-2002-bleu} and COMET~\cite{rei2020cometneuralframeworkmt} remain essential for measuring linguistic adequacy, especially for S2TT. However, for S2ST and streaming ST, translation quality is only one part of system behavior. Evaluation must also account for how translated speech is realized, what speaker and expressive information is preserved, and how outputs are timed during generation.

Relevant evaluation practices have therefore emerged across several lines of work. S2ST studies often include speech-side measures such as speech quality, intelligibility, speaker similarity, or expressiveness, while simultaneous translation toolkits focus on latency and streaming behavior~\cite{ma-etal-2020-simuleval}. Prior work has also shown that evaluating S2ST outputs with automatic metrics remains challenging~\cite{salesky2021assessingevaluationmetricsspeechtospeech}. However, these evaluation dimensions are usually applied under task-specific protocols, making it difficult to systematically compare S2TT, S2ST, offline, and streaming systems within a common framework.

To address this gap, we present OpenSTBench, a multidimensional evaluation framework for speech translation. In OpenSTBench, \emph{unified} refers to evaluating heterogeneous S2TT and S2ST systems through a shared sample record, a common evaluator interface, and a consistent output schema, with metrics selected according to the available output modality and timing information. OpenSTBench integrates translation quality, speech quality, and temporal quality while supporting both offline and streaming settings.

Using OpenSTBench, we conduct a comparative study of representative speech translation systems and show that system rankings vary substantially across dimensions. This suggests that speech translation systems should be compared and selected with respect to application priorities and cross-dimensional trade-offs rather than a single global ranking.

Our contributions are as follows:

\begin{itemize}
    \item We present OpenSTBench, a unified framework for multidimensional speech translation evaluation that consolidates translation quality, speech quality, and temporal quality across S2TT, S2ST, offline, and streaming settings.

    \item We establish a reproducible evaluation protocol that standardizes data organization, scoring, and aggregation across heterogeneous ST systems.

    \item We will release OpenSTBench as an extensible open-source Python package with a shared input format and modular evaluators, enabling users to evaluate their own system outputs, obtain multidimensional reports, and adapt the framework to new datasets, language pairs, and research goals.

    \item We provide empirical evidence that representative speech translation systems exhibit substantial cross-dimensional ranking variation, motivating application-oriented system selection rather than a winner-takes-all ranking.
\end{itemize}

\section{Related Work}

\subsection{Evaluation in Speech Translation}

Evaluation in speech translation (ST) has traditionally followed machine translation practice for measuring translation quality, with BLEU and COMET widely used to compare system outputs against reference translations, especially in S2TT. As ST systems increasingly include S2ST and streaming generation, evaluation has also begun to incorporate speech-side and timing-related aspects such as speech naturalness, intelligibility, speaker similarity, and latency. However, these dimensions are often evaluated separately and under task-specific protocols. This motivates our focus on a systematic evaluation framework that jointly organizes translation quality, speech quality, and temporal quality, including temporal consistency and latency, for heterogeneous ST systems.

\subsection{Speech-specific and Streaming Dimensions}

% Several evaluation dimensions relevant to ST have been developed in neighboring speech-related fields. These include speech quality assessment, speaker preservation, emotion and paralinguistic evaluation, and latency- or timing-related measures for streaming and simultaneous settings. Together, these lines of work provide many of the ingredients needed for broader ST evaluation, but they are typically applied in isolation or within task-specific experimental setups.

Several representative works in neighboring speech fields instantiate evaluation dimensions relevant to ST. UTMOS~\cite{utmos} predicts MOS-style judgments for speech naturalness and quality. Resemblyzer, following the GE2E/d-vector speaker verification paradigm~\cite{GE2E}, and WavLM~\cite{WavLM} support embedding-based speaker similarity, while emotion2vec~\cite{emotion2vec} provides affective representations. Acoustic event detection has been studied in audio benchmarks such as AudioSet~\cite{audioset} and DCASE~\cite{dcase}, but not for event preservation in cross-lingual speech translation. For streaming and simultaneous settings, SimulEval~\cite{ma-etal-2020-simuleval} standardizes quality-latency evaluation. These works provide useful ingredients, but are designed for separate evaluation settings.

%这里是否需要引用几篇文章来说明一下

%Recent work in NLP and speech has increasingly emphasized unified benchmarks, common evaluation protocols, and multi-metric analysis. Our work follows this direction. OpenSTBench does not introduce a new standalone metric; instead, it organizes translation quality, speech quality, speaker preservation, emotion and paralinguistic fidelity, temporal consistency, and streaming latency within a unified evaluation framework for speech translation.
% figure/openstbench_unification.tex
\begin{table}[t]
\centering
\scriptsize
\setlength{\tabcolsep}{3pt}
\begin{tabular}{lccc}
\toprule
Resource
& Translation Quality
& Speech Quality
& Temporal Quality \\
\midrule
SimulEval
& \textcolor[HTML]{2E8B57}{\checkmark}
& \textcolor[HTML]{C0392B}{\ding{55}}
& \textcolor[HTML]{2E8B57}{\checkmark} \\

\textbf{OpenSTBench}
& \textcolor[HTML]{2E8B57}{\checkmark}
& \textcolor[HTML]{2E8B57}{\checkmark}
& \textcolor[HTML]{2E8B57}{\checkmark} \\
\bottomrule
\end{tabular}
\caption{Comparison with SimulEval.}
\label{tab:benchmark_comparison}
\end{table}

\begin{figure}[t]
    \centering
    \includegraphics[width=\columnwidth]{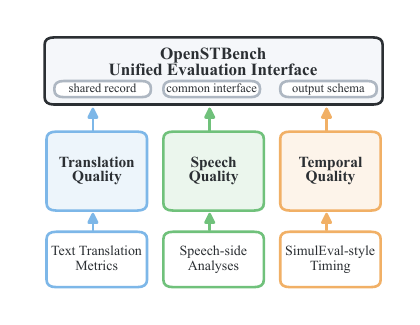}
    \caption{Conceptual positioning of OpenSTBench.}
    \label{fig:openstbench_unification}
\end{figure}

Table~\ref{tab:benchmark_comparison} compares OpenSTBench with SimulEval, a representative toolkit for simultaneous text and speech translation evaluation. SimulEval standardizes online interaction and quality-latency measurement, which OpenSTBench builds on for temporal evaluation. OpenSTBench further organizes text translation metrics, speech-side analyses, and timing evaluation tools under a shared record, common interface, and output schema, enabling unified evaluation of translation quality, speech quality, and temporal quality for heterogeneous ST systems, as illustrated in Figure~\ref{fig:openstbench_unification}.
\section{OpenSTBench}
\label{sec:openstbench}

\begin{figure*}[t]
    \centering
    \includegraphics[width=\textwidth]{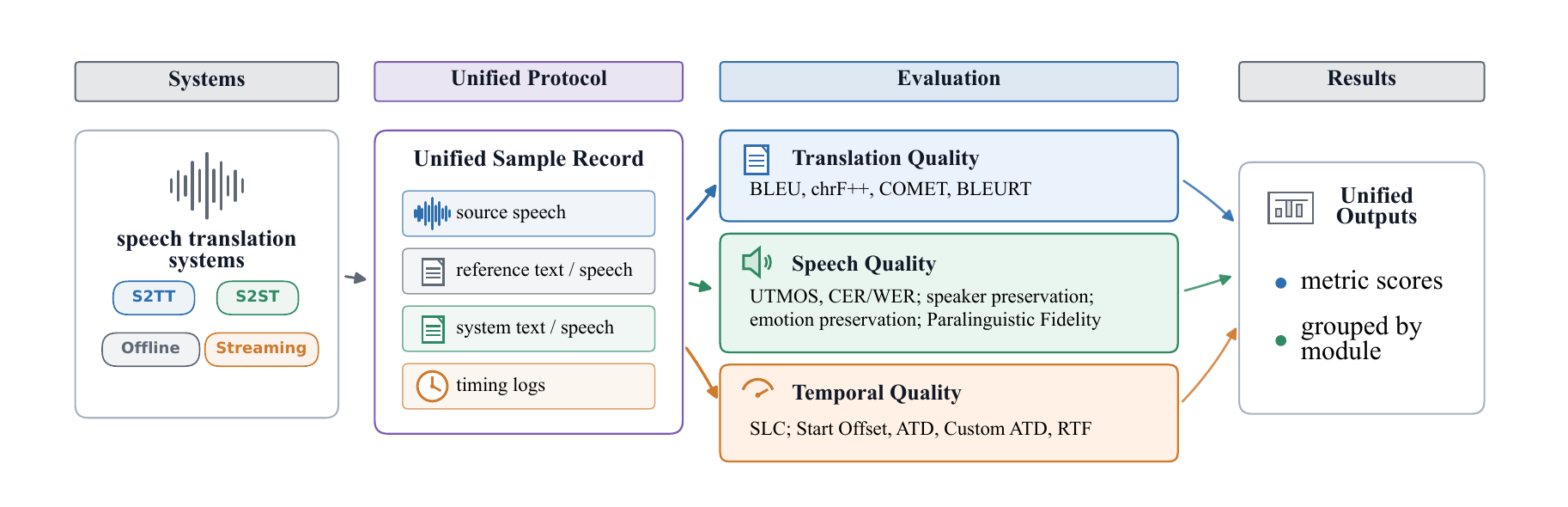}
    \caption{Overview of OpenSTBench. The framework represents heterogeneous S2TT and S2ST outputs with a shared evaluation record and reports comparable profiles across translation quality, speech quality, and temporal quality.}
    \label{fig:openstbench_overview}
\end{figure*}

\subsection{Design Goals}

Figure~\ref{fig:openstbench_overview} illustrates the workflow of OpenSTBench. The framework provides a common protocol for evaluating S2TT and S2ST systems in both offline and streaming settings. It organizes translation quality, speech quality, and temporal quality under a shared input format, evaluator interface, and output schema, supporting reproducible scoring and modular extension to new datasets and evaluation components.

\subsection{Evaluation Dimensions}

OpenSTBench organizes evaluation into three groups: translation quality, speech quality, and temporal quality. Table~\ref{tab:metric_overview} summarizes the corresponding metrics, applicable system types, and score directions.

\begin{table}[t]
\centering
\scriptsize
\setlength{\tabcolsep}{3pt}
\begin{tabular}{lp{0.47\columnwidth}l}
\toprule
Dimension & Metrics & Applies to \\
\midrule
\multicolumn{3}{l}{\textit{Translation Quality}} \\
Translation quality
& BLEU, chrF++, COMET, BLEURT
& S2TT, S2ST \\

\midrule
\multicolumn{3}{l}{\textit{Speech Quality}} \\
Speech quality
& UTMOS, CER/WER
& S2ST \\
Speaker preservation
& Resemblyzer, WavLM
& S2ST \\
Emotion preservation
& Emotion2Vec, Emotion Acc.
& S2ST \\
Paralinguistic fidelity
& Event Content F1, Event Timing F1
& S2ST \\

\midrule
\multicolumn{3}{l}{\textit{Temporal Quality}} \\
Temporal consistency
& SLC 0.2, SLC 0.4
& S2ST \\
Streaming latency
& Start Offset, ATD, Custom ATD
& Streaming \\
Offline efficiency
& RTF
& Offline local \\
\bottomrule
\end{tabular}
\caption{Evaluation dimensions and metrics in OpenSTBench. Applies-to indicates the system types for which each metric group is computed.}
\label{tab:metric_overview}
\end{table}

\paragraph{Translation quality.}
Translation quality measures whether the system conveys the linguistic content of the source speech correctly. OpenSTBench evaluates this dimension with sacreBLEU~\cite{post-2018-sacreBLEU}, chrF++~\cite{popovic-2017-chrf++}, COMET\footnote{
\url{https://huggingface.co/Unbabel/wmt22-comet-da}}~\cite{rei2020cometneuralframeworkmt}, and BLEURT\footnote{PyTorch implementation from \url{https://github.com/lucadiliello/bleurt-pytorch} with the checkpoint from \url{https://huggingface.co/lucadiliello/BLEURT-20}.}~\cite{sellam-etal-2020-bleurt}. These metrics compare system outputs with reference translations and capture complementary lexical, character-level, and semantic aspects of translation quality. This dimension applies to both S2TT and S2ST systems, since target speech can be evaluated through its associated generated text or transcription when needed.

\paragraph{Speech quality.}
Speech quality evaluates whether target speech is natural, faithfully realizes the intended text, and preserves relevant speech-side attributes. OpenSTBench measures naturalness with UTMOS\footnote{\url{https://github.com/tarepan/SpeechMOS}} and realization fidelity with ASR-based error rates: generated speech is transcribed with Whisper\footnote{\url{https://huggingface.co/openai/whisper-medium}} and compared with the generated target text using CER for Chinese, Japanese, and Korean targets and WER otherwise, where lower scores indicate better realization.

For speaker preservation, OpenSTBench compares generated speech with target-language reference speech using WavLM\footnote{
\url{https://github.com/microsoft/UniSpeech/tree/main/downstreams/speaker_verification}}~\cite{WavLM} and Resemblyzer\footnote{\url{https://github.com/resemble-ai/Resemblyzer}} speaker-similarity evaluators. For emotion preservation, it reports Emotion2Vec-based cosine similarity\footnote{\url{https://modelscope.cn/models/iic/emotion2vec_plus_large}} and audio emotion classification accuracy. For paralinguistic fidelity, it reports Event Content F1 and Event Timing F1 based on acoustic events detected with CLAP\footnote{\url{https://huggingface.co/laion/clap-htsat-fused}}, measuring event-type/count preservation and relative event-timing preservation, respectively.

\paragraph{Temporal Quality.}
Temporal quality includes temporal consistency for generated speech and latency for streaming systems. Temporal consistency measures whether the generated speech preserves the coarse duration structure of the source speech. Following duration-consistency evaluation in expressive S2ST~\cite{cheng2025uniss, wu2023videodubber}, OpenSTBench reports Speech Length Compliant (SLC) score. To evaluate speech length control performance, \(\mathrm{SLC}_p\) refers to the percentage of samples whose \(\mathrm{ratio} \in [1-p, 1+p]\), where the ratio is defined as
\begin{equation}
\mathrm{ratio} =
\frac{\sum_{i=0}^{T_{y'}} d_i}
{\sum_{j=0}^{T_x} d_j}.
\label{eq:ratio}
\end{equation}
Here, \(T_{y'}\) and \(T_x\) denote the translated and source speech token sequences, and \(d_i\) and \(d_j\) denote token durations. We report \(\mathrm{SLC}_{0.2}\) and \(\mathrm{SLC}_{0.4}\). Higher values indicate better temporal consistency.

Latency measures the responsiveness of streaming systems. OpenSTBench reports Start Offset, Average Token Delay (ATD), Custom ATD, and Real-Time Factor (RTF), with computation-aware variants when applicable. The implementation follows the evaluation interface of SimulEval~\cite{ma-etal-2020-simuleval}, while extending the scoring pipeline to support speech-to-speech outputs. Start Offset measures how long the system waits before emitting output. ATD measures content-level delay between the source and aligned generated output. For speech outputs, ATD is computed through aligned textual representations of the generated speech. Custom ATD subtracts target-audio playback duration from the measured delay to better isolate generation-side latency. RTF measures processing time divided by source audio duration.

\subsection{Unified Protocol and Open-source Reuse}

For open-source reuse, OpenSTBench separates system outputs from evaluation modules: each system is represented with the same input schema, while evaluators are invoked according to the available output modality and timing information. This design allows new systems to be compared under the same reporting format and allows individual evaluation modules to be replaced or extended without changing the overall protocol.

\section{Experimental Setup}

\subsection{Compared Systems}

We evaluate representative speech translation systems in both streaming and offline settings.

For streaming settings, we evaluate Qwen3-LiveTranslate\footnote{\url{https://www.alibabacloud.com/help/en/model-studio/qwen3-livetranslate-flash-realtime}}, Doubao AST 2.0\footnote{\url{https://www.volcengine.com/docs/4640/127504}}, GPT Realtime Translate\footnote{\url{https://developers.openai.com/api/docs/models/gpt-realtime-translate}}, and Baidu Realtime ST\footnote{\url{https://cloud.baidu.com/doc/MT/s/Sl9p2h5k9}}. Qwen3-LiveTranslate, Doubao AST 2.0, and GPT Realtime Translate generate target speech and are evaluated as streaming S2ST systems. Baidu Realtime ST receives streaming audio input but emits sentence-final text output, and is therefore treated as a streaming-input S2TT baseline rather than a fully incremental S2ST system.

For offline settings, we evaluate SeamlessM4T-v2-Large\footnote{\url{https://huggingface.co/facebook/seamless-m4t-v2-large}}~\cite{communication2023seamlessmultilingualexpressivestreaming} and UniSS\footnote{\url{https://huggingface.co/cmots/UniSS}}~\cite{cheng2025uniss}. Both receive complete speech segments as input and generate target speech, and are therefore evaluated on translation quality and all speech-side dimensions except streaming latency. They correspond to 2.3B and 1.5B parameters, respectively; parameter counts are not publicly disclosed for the API-based streaming systems.

\subsection{Datasets}

We use separate datasets for different evaluation dimensions. Public datasets are used for translation quality, speech quality, emotion preservation, paralinguistic fidelity, temporal consistency, and latency, while a LibriTTS-based paired speaker set is constructed for speaker preservation.
Table~\ref{tab:dataset_overview} summarizes the datasets used for each evaluation target.

\begin{table}[t]
\centering
\scriptsize
\setlength{\tabcolsep}{3pt}
\begin{tabular}{llcl}
\toprule
Dataset & Direction & \#Samples & Evaluation aspect \\
\midrule
MSLT dev & EN$\rightarrow$ZH & 1,000 & TQ, SQ, TC, latency \\
MSLT dev & ZH$\rightarrow$EN & 1,000 & TQ, SQ, TC, latency \\
LibriTTS paired set & EN$\rightarrow$ZH & 300 & Speaker preservation \\
LibriTTS paired set & ZH$\rightarrow$EN & 300 & Speaker preservation \\
RAVDESS & EN$\rightarrow$ZH & 1,440 & Emotion preservation \\
MCAE-SPPS & ZH$\rightarrow$EN & 1,029 & Emotion preservation \\
NonverbalTTS test & EN$\rightarrow$ZH & 359 & Paralinguistic fidelity \\
SynParaSpeech & ZH$\rightarrow$EN & 500 & Paralinguistic fidelity \\
\bottomrule
\end{tabular}
\caption{Datasets used in OpenSTBench. TQ denotes translation quality, SQ denotes speech quality, and TC denotes temporal consistency. \#Samples denotes the number of samples evaluated in the main experiments.}
\label{tab:dataset_overview}
\end{table}

\textbf{MSLT for general ST evaluation.}
We use MSLT~\cite{MSLT} for translation quality, speech quality, temporal consistency, and latency evaluation in both EN$\rightarrow$ZH and ZH$\rightarrow$EN. In the main experiments, we use the dev split and cap evaluation at 1,000 samples per direction.

\label{sec:speaker_anchor}
\textbf{LibriTTS-based paired speaker set.}
Speaker preservation requires a same-language speaker anchor, since cross-language comparison can conflate speaker identity with language mismatch. We verify this with same-speaker LibriTTS pairs synthesized under English--English and English--Chinese conditions. As shown in Table~\ref{tab:speaker_similarity_anchor_analysis}, WavLM similarity drops substantially in the cross-language condition, motivating same-language comparison.

We construct a 300-sample paired speaker set by sampling LibriTTS utterances from 35 speakers~\cite{libritts}. Source transcripts are translated into Chinese with \texttt{qwen3-235b-a22b-instruct-2507}, and Chinese reference speech is synthesized with Qwen3-TTS~\cite{qwen3tts} using the source utterance as the speaker prompt. For speaker-preservation evaluation, EN$\rightarrow$ZH outputs are compared with synthesized Chinese reference speech, while ZH$\rightarrow$EN outputs are compared with the original LibriTTS speech.

\begin{table}[t]
\centering
\begin{tabular}{lcc}
\toprule
Anchor Pair & Resemblyzer$\uparrow$ & WavLM$\uparrow$ \\
\midrule
EN--EN & 0.9073 & 0.7721 \\
EN--ZH & 0.8462 & 0.4556 \\
\bottomrule
\end{tabular}

\caption{Same-speaker speaker-similarity diagnostic on the LibriTTS-based paired speaker set. English--English compares original LibriTTS speech with Qwen3-TTS synthesized English speech, while English--Chinese compares original LibriTTS speech with Qwen3-TTS synthesized Chinese speech under the same speaker prompt. Higher is better.}
\label{tab:speaker_similarity_anchor_analysis}
\end{table}

\begin{figure*}[t]
    \centering
    \raisebox{0.2ex}{\textcolor[HTML]{7DB7E8}{\rule{1.6em}{1.0pt}}} Translation Quality
    \quad
    \raisebox{0.2ex}{\textcolor[HTML]{70C17B}{\rule{1.6em}{1.0pt}}} Speech Quality
    \quad
    \raisebox{0.2ex}{\textcolor[HTML]{F1B066}{\rule{1.6em}{1.0pt}}} Temporal Quality

    \vspace{0.4em}

    \begin{minipage}{0.49\textwidth}
        \centering
        \includegraphics[width=\linewidth]{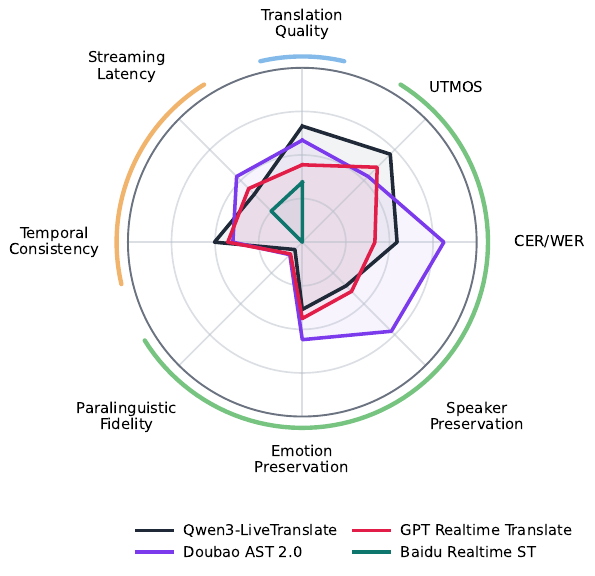}\\[-0.2em]
        \textbf{(a) Streaming Systems}
    \end{minipage}
    \hfill
    \begin{minipage}{0.49\textwidth}
        \centering
        \includegraphics[width=\linewidth]{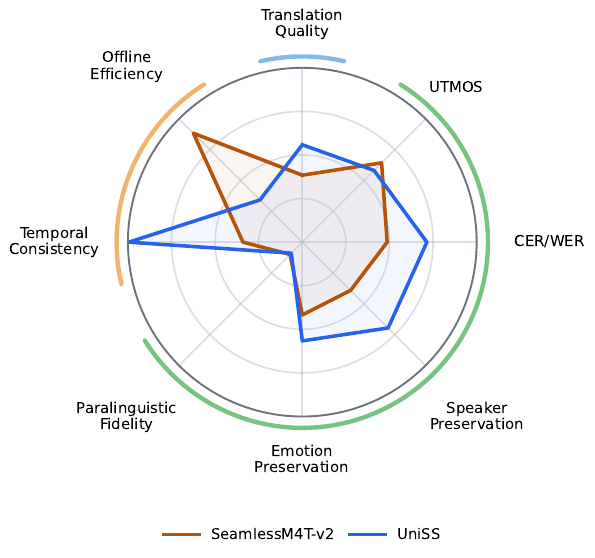}\\[-0.2em]
        \textbf{(b) Offline Systems}
    \end{minipage}

    \caption{Bilingual-average radar plots of normalized cross-dimensional scores for streaming and offline systems. Axes are grouped into translation quality, speech quality, and temporal quality. Streaming latency is shown for streaming systems, whereas offline efficiency is shown for offline systems. Missing target-speech dimensions for text-only systems are assigned a normalized score of 0.}
    \label{fig:radar}
\end{figure*}

\textbf{RAVDESS and MCAE-SPPS for emotion preservation.}
We use RAVDESS~\cite{ravdess} for EN$\rightarrow$ZH and MCAE-SPPS~\cite{mcae-spss} for ZH$\rightarrow$EN emotion preservation. We evaluate all 1,440 valid RAVDESS utterances and use class-balanced sampling on MCAE-SPPS with a fixed random seed, yielding 1,029 evaluation samples.

\textbf{NonverbalTTS and SynParaSpeech for paralinguistic fidelity.}
We use NonverbalTTS~\cite{nonverbaltts} for EN$\rightarrow$ZH and SynParaSpeech~\cite{synparaspeech} for ZH$\rightarrow$EN paralinguistic fidelity. We evaluate the 359-sample NonverbalTTS test split and sample 500 SynParaSpeech instances with a fixed random seed. Source-side event locations are approximate rather than manually annotated frame-level timestamps. For NonverbalTTS, we estimate event locations by mapping the positions of nonverbal tags in text to utterance duration. For SynParaSpeech, we derive event locations by aligning text-level event tags with VAD-based segment timing.

\subsection{Metrics}

We use the OpenSTBench metrics described in Section~\ref{sec:openstbench}, selected according to each system's output modality and timing information. All systems are evaluated for translation quality with sacreBLEU, chrF++, COMET, and BLEURT. Systems with target speech are additionally evaluated with UTMOS, CER/WER, speaker preservation, emotion preservation, paralinguistic fidelity, and temporal consistency. Temporal quality is evaluated with Start Offset, ATD, Custom ATD, and RTF when applicable. Additional implementation choices and evaluation settings are summarized in Appendix~\ref{app:evaluation_settings}.

\subsection{Evaluation Protocol}
\label{sec:evaluation_protocol}

For streaming systems, we follow each system's native online interface and feed source speech incrementally using the interface's input granularity to reduce segmentation-induced latency bias. For streaming S2ST systems, we retain generated speech and synchronized text; for S2TT systems, we retain text outputs and compute only applicable metrics. Since Baidu Realtime ST emits sentence-final text, its latency is reported as sentence-final delay rather than incremental S2ST latency.

For offline S2ST systems, we run inference on complete speech segments and report RTF rather than incremental delay metrics.

\subsection{Implementation Details}

We use OpenSTBench for post-processing, alignment, and scoring. Translation outputs are sentence-aligned with references; speech-side metrics are computed on generated audio for S2ST systems, while S2TT systems use only applicable text and latency metrics.

Speaker preservation uses the same-language references from the LibriTTS-based paired speaker set. For paralinguistic fidelity, source events come from dataset annotations, and target events are detected and localized from generated speech with CLAP under the same candidate label set.

Offline RTF is measured on a single NVIDIA GeForce RTX 3090 GPU with each model's native inference configuration. UniSS follows its official VAD-based chunked pipeline, while SeamlessM4T-v2-Large uses full-utterance generation. We do not report API-system RTF, model size, or service-side compute budget because provider-side runtime, scheduling, hardware, and parameter counts are not publicly exposed.

We use official or default inference configurations for locally runnable systems and do not perform hyperparameter search.
\section{Results and Analysis}

\begin{table*}[tb]
\centering
\scriptsize
\setlength{\tabcolsep}{4pt}
\begin{tabular}{lcccccccc}
\toprule
& \multicolumn{4}{c}{EN$\rightarrow$ZH} & \multicolumn{4}{c}{ZH$\rightarrow$EN} \\
\cmidrule(lr){2-5} \cmidrule(lr){6-9}
Model & \multicolumn{4}{c}{Translation Quality} & \multicolumn{4}{c}{Translation Quality} \\
\cmidrule(lr){2-5} \cmidrule(lr){6-9}
& BLEU$\uparrow$ & chrF++$\uparrow$ & COMET$\uparrow$ & BLEURT$\uparrow$
& BLEU$\uparrow$ & chrF++$\uparrow$ & COMET$\uparrow$ & BLEURT$\uparrow$ \\
\midrule
\multicolumn{9}{l}{\textit{Streaming Models}} \\
Qwen3-LiveTranslate
& \textbf{43.2726} & \textbf{29.4047} & \textbf{0.8626} & \textbf{0.7058}
& \textbf{24.6421} & \textbf{48.0138} & \textbf{0.7041} & \textbf{0.5351} \\
Doubao AST 2.0
& \underline{36.7764} & \underline{24.9965} & \underline{0.8482} & \underline{0.6750}
& \underline{19.3916} & 41.7650 & \underline{0.6935} & 0.5055 \\
GPT Realtime Translate
& 21.4059 & 14.2556 & 0.7870 & 0.5816
& 16.1659 & 39.8521 & 0.6606 & 0.4904 \\
Baidu Realtime ST
& 16.1149 & 14.5125 & 0.7470 & 0.5754
& 8.9036 & 24.6700 & 0.6181 & 0.4637 \\
\midrule
\multicolumn{9}{l}{\textit{Offline Models}} \\
SeamlessM4T-v2
& 23.5334 & 15.8207 & 0.7207 & 0.5070
& 14.5280 & 36.1125 & 0.6378 & 0.4857 \\
UniSS
& 34.1008 & 23.9559 & 0.8104 & 0.6394
& 18.7520 & \underline{44.5369} & 0.6863 & \underline{0.5203} \\
\bottomrule
\end{tabular}

\caption{Translation quality results for EN$\rightarrow$ZH and ZH$\rightarrow$EN. BLEU denotes sacreBLEU. Higher is better for all metrics. Best and second-best results are bolded and underlined, respectively.}
\label{tab:translation_quality}
\end{table*}

\begin{table*}[tb]
\centering
\begin{subtable}[t]{\textwidth}
\centering
\caption{EN$\rightarrow$ZH}
\scriptsize
\setlength{\tabcolsep}{4pt}
\begin{tabular}{lcccccccc}
\toprule
& \multicolumn{2}{c}{Speech Quality} & \multicolumn{2}{c}{Speaker Preservation} & \multicolumn{2}{c}{Emotion} & \multicolumn{2}{c}{Paralinguistic Fidelity} \\
\cmidrule(lr){2-3} \cmidrule(lr){4-5} \cmidrule(lr){6-7} \cmidrule(lr){8-9}
Model & UTMOS$\uparrow$ & CER$\downarrow$ & Resemblyzer$\uparrow$ & WavLM$\uparrow$ & E2V$\uparrow$ & Acc.$\uparrow$ & Event Content F1$\uparrow$ & Event Timing F1$\uparrow$ \\
\midrule
\multicolumn{9}{l}{\textit{Streaming Models}} \\
Qwen3-LiveTranslate        & \underline{3.6028} & 0.1371 & 0.5967 & 0.2255 & 0.6280 & 0.1056 & 0.0570 & 0.0303 \\
Doubao AST 2.0      & 2.8675 & \textbf{0.0782} & \underline{0.8261} & \textbf{0.6849} & \underline{0.7247} & \textbf{0.2660} & \textbf{0.1492} & \textbf{0.0843} \\
GPT Realtime Translate & 3.1749 & 0.2729 & 0.5767 & 0.2275 & 0.6481 & 0.1000 & 0.0967 & 0.0686 \\
\midrule
\multicolumn{9}{l}{\textit{Offline Models}} \\
SeamlessM4T-v2  & \textbf{3.7160} & 0.1685 & 0.6086 & 0.2669 & 0.6323 & 0.0840 & \underline{0.1279} & 0.0706 \\
UniSS           & 3.2409 & \underline{0.1080} & \textbf{0.8468} & \underline{0.6291} & \textbf{0.7383} & \underline{0.2625} & 0.1070 & \underline{0.0753} \\
\bottomrule
\end{tabular}
\end{subtable}

\vspace{0.6em}

\begin{subtable}[t]{\textwidth}
\centering
\caption{ZH$\rightarrow$EN}
\scriptsize
\setlength{\tabcolsep}{4pt}
\begin{tabular}{lcccccccc}
\toprule
& \multicolumn{2}{c}{Speech Quality} & \multicolumn{2}{c}{Speaker Preservation} & \multicolumn{2}{c}{Emotion} & \multicolumn{2}{c}{Paralinguistic Fidelity} \\
\cmidrule(lr){2-3} \cmidrule(lr){4-5} \cmidrule(lr){6-7} \cmidrule(lr){8-9}
Model & UTMOS$\uparrow$ & WER$\downarrow$ & Resemblyzer$\uparrow$ & WavLM$\uparrow$ & E2V$\uparrow$ & Acc.$\uparrow$ & Event Content F1$\uparrow$ & Event Timing F1$\uparrow$ \\
\midrule
\multicolumn{9}{l}{\textit{Streaming Models}} \\
Qwen3-LiveTranslate        & \textbf{4.1054} & 0.1368 & 0.5670 & 0.0288 & 0.6807 & 0.1297 & 0.1175 & 0.0349 \\
Doubao AST 2.0      & 3.3993 & \textbf{0.0346} & \underline{0.8397} & \textbf{0.5392} & \underline{0.8563} & \textbf{0.3897} & 0.1322 & 0.0398 \\
GPT Realtime Translate & \underline{3.6848} & 0.0773 & 0.6389 & 0.1541 & 0.8309 & 0.1728 & \underline{0.1346} & \textbf{0.0814} \\
\midrule
\multicolumn{9}{l}{\textit{Offline Models}} \\
SeamlessM4T-v2  & 3.4288 & 0.1395 & 0.5976 & 0.0921 & 0.8094 & 0.1438 & \textbf{0.1502} & 0.0485 \\
UniSS           & 3.4139 & \underline{0.0632} & \textbf{0.8459} & \underline{0.4619} & \textbf{0.9035} & \underline{0.3608} & 0.1243 & \underline{0.0496} \\
\bottomrule
\end{tabular}
\end{subtable}

\caption{Speech quality results for EN$\rightarrow$ZH and ZH$\rightarrow$EN. Lower is better for CER/WER, and higher is better otherwise. Baidu Realtime ST is omitted because these metrics require target speech. Best and second-best results are bolded and underlined, respectively.}
\label{tab:speech_attributes}
\end{table*}

\begin{table*}[tb]
\centering
\begin{subtable}[t]{\textwidth}
\centering
\caption{EN$\rightarrow$ZH}
\scriptsize
\setlength{\tabcolsep}{3pt}
\begin{tabular}{lcccccc}
\toprule
& \multicolumn{4}{c}{Latency} & \multicolumn{2}{c}{Temporal Consistency} \\
\cmidrule(lr){2-5} \cmidrule(lr){6-7}
Model & Start Offset (ms)$\downarrow$ & ATD (ms)$\downarrow$ & Custom ATD (ms)$\downarrow$ & RTF$\downarrow$ & SLC 0.2$\uparrow$ & SLC 0.4$\uparrow$ \\
\midrule
\multicolumn{7}{l}{\textit{Streaming Models}} \\
Qwen3-LiveTranslate
& 3656.78 & 5138.88 & \underline{3446.17} & -- & \underline{0.4209} & \underline{0.7500} \\
Doubao AST 2.0
& \textbf{2320.27} & \textbf{4072.74} & \textbf{2977.44} & -- & 0.2725 & 0.6808 \\
GPT Realtime Translate
& \underline{2696.76} & 5956.45 & 3480.21 & -- & 0.2023 & 0.4162 \\
Baidu Realtime ST
& 4960.83 & \underline{4733.85} & 4733.85 & -- & -- & -- \\
\midrule
\multicolumn{7}{l}{\textit{Offline Models}} \\
SeamlessM4T-v2
& -- & -- & -- & \textbf{0.3010} & 0.4111 & 0.7333 \\
UniSS
& -- & -- & -- & \underline{1.5449} & \textbf{0.9940} & \textbf{0.9980} \\
\bottomrule
\end{tabular}
\end{subtable}

\vspace{0.6em}

\begin{subtable}[t]{\textwidth}
\centering
\caption{ZH$\rightarrow$EN}
\scriptsize
\setlength{\tabcolsep}{3pt}
\begin{tabular}{lcccccc}
\toprule
& \multicolumn{4}{c}{Latency} & \multicolumn{2}{c}{Temporal Consistency} \\
\cmidrule(lr){2-5} \cmidrule(lr){6-7}
Model & Start Offset (ms)$\downarrow$ & ATD (ms)$\downarrow$ & Custom ATD (ms)$\downarrow$ & RTF$\downarrow$ & SLC 0.2$\uparrow$ & SLC 0.4$\uparrow$ \\
\midrule
\multicolumn{7}{l}{\textit{Streaming Models}} \\
Qwen3-LiveTranslate
& 4730.69 & 7204.57 & 5153.51 & -- & 0.2670 & 0.5764 \\
Doubao AST 2.0
& \underline{3163.02} & \textbf{5526.35} & \textbf{4368.24} & -- & 0.1527 & 0.4867 \\
GPT Realtime Translate
& \textbf{3019.25} & 7220.77 & \underline{4755.51} & -- & \underline{0.4521} & \underline{0.6383} \\
Baidu Realtime ST
& 7623.02 & \underline{6926.38} & 6926.38 & -- & -- & -- \\
\midrule
\multicolumn{7}{l}{\textit{Offline Models}} \\
SeamlessM4T-v2
& -- & -- & -- & \textbf{0.1648} & 0.0240 & 0.1920 \\
UniSS
& -- & -- & -- & \underline{1.0838} & \textbf{0.9919} & \textbf{0.9960} \\
\bottomrule
\end{tabular}
\end{subtable}

\caption{Temporal quality results for EN$\rightarrow$ZH and ZH$\rightarrow$EN. Start Offset, ATD, and Custom ATD are reported for streaming systems, while RTF is reported for offline local models. SLC denotes Speech Length Compliant score. Lower is better for latency and RTF, and higher is better for SLC. Baidu Realtime ST has no SLC because SLC requires target speech. Best and second-best results are bolded and underlined, respectively.}
\label{tab:latency_temporal}
\end{table*}

We report results from three complementary perspectives: translation quality, speech quality, and temporal quality. Rather than deriving a single leaderboard, we characterize how systems differ across dimensions. Unless otherwise specified, table entries report aggregate scores over the evaluated samples for each language direction, computed once for each system. Metrics defined at the sample level are averaged over samples. Figure~\ref{fig:radar} summarizes these profiles with bilingual averages for streaming and offline systems.

Figure~\ref{fig:radar} uses fixed-range normalization. For each metric value \(x\), we compute
\begin{equation}
s(x)=\mathrm{clip}\!\left(\frac{x-a}{b-a},\,0,\,1\right)
\end{equation}
when higher is better, and
\begin{equation}
s(x)=\mathrm{clip}\!\left(\frac{b-x}{b-a},\,0,\,1\right)
\end{equation}
when lower is better, where \([a,b]\) is fixed across systems and language directions. All radar scores lie in \([0,1]\), with larger values indicating better performance; the fixed ranges are listed in Appendix~\ref{app:radar_ranges}. Axis scores are averaged over normalized metrics within each direction and then over EN$\rightarrow$ZH and ZH$\rightarrow$EN. Multi-metric axes are translation quality (BLEU, chrF++, COMET, BLEURT), speaker preservation (Resemblyzer, WavLM), emotion preservation (E2V, emotion classification accuracy), paralinguistic fidelity (Event Content F1, Event Timing F1), temporal consistency (SLC 0.2, SLC 0.4), and streaming latency (Start Offset, ATD, Custom ATD). UTMOS, CER/WER, and offline efficiency are standalone axes, with offline efficiency computed from RTF.

\subsection{Translation Quality}

We first examine translation quality, which remains the standard comparison axis in speech translation evaluation.

Table~\ref{tab:translation_quality} shows that text-side metrics produce the clearest ranking among all evaluation blocks. Qwen3-LiveTranslate is consistently strongest across both directions and metric families, indicating robust linguistic fidelity. Among the remaining systems, Doubao AST 2.0 and UniSS form the strongest alternatives in streaming and offline settings, respectively, while GPT Realtime Translate and Baidu Realtime ST are weaker on text-side quality.

This ranking is useful but incomplete. As the following sections show, systems with strong translation quality do not necessarily preserve speech-side attributes or time-related behavior equally well.

\subsection{Speech Quality}

We evaluate speech naturalness, ASR-based CER/WER, speaker preservation, emotion preservation, and paralinguistic fidelity.

Table~\ref{tab:speech_attributes} shows patterns different from translation quality. UTMOS and ASR-based CER/WER do not always favor the same systems, indicating that perceived speech naturalness and accurate realization of the generated target text should be evaluated separately. UniSS and Doubao AST 2.0 show consistently strong speaker preservation, while emotion preservation varies between embedding-based similarity and classification accuracy, supporting the use of complementary preservation evaluators.

Paralinguistic fidelity remains weak across systems, with low absolute Event Content F1 and Event Timing F1 despite direction-dependent ranking differences. This indicates that preserving acoustic event content and timing remains a major challenge for current speech translation systems.

\subsection{Temporal Quality}

Finally, we examine temporal quality, including latency and temporal consistency.

Table~\ref{tab:latency_temporal} shows that responsiveness, temporal consistency, and offline efficiency capture distinct properties. Streaming delay metrics favor systems that begin and update output earlier, but these systems are not necessarily the most temporally consistent. Conversely, offline systems can be efficient in RTF while still differing substantially in SLC.

Baidu Realtime ST also illustrates an important evaluation distinction: although it accepts streaming input, it emits sentence-final text output and therefore occupies a different operating point from fully incremental S2ST systems. OpenSTBench makes this distinction explicit by reporting only applicable metrics for each output regime.

These results show that time-related behavior cannot be summarized by a single latency or efficiency number. Streaming responsiveness, duration preservation, and offline computation each reflect different deployment-relevant constraints.

\subsection{Overall Comparison}

Across all result blocks, no single system dominates every dimension: strong text-side performance does not necessarily imply strong speech-side preservation, and low latency does not guarantee temporal consistency. These cross-dimensional differences show that speech translation systems cannot be fully characterized by text-side quality alone, motivating application-oriented comparison under a unified multidimensional protocol.
\section{Conclusion}

We presented OpenSTBench, a unified framework for multidimensional speech translation evaluation across S2TT, S2ST, offline, and streaming settings. By organizing translation quality, speech quality, and temporal quality under a shared protocol, OpenSTBench enables comparison across heterogeneous system outputs. Experiments on representative systems show substantial cross-dimensional variation, motivating application-oriented comparison rather than a single global ranking. As an open-source and extensible framework, OpenSTBench provides a reproducible basis for future speech translation evaluation.
\section*{Limitations}

OpenSTBench has several limitations. First, the current experiments focus on EN$\leftrightarrow$ZH, so broader empirical validation across additional language pairs remains for future work. Second, several metrics for speech quality and preservation rely on automatic evaluators, whose correspondence to human judgments still requires further validation. Third, streaming S2ST, streaming S2TT, and offline S2ST systems are not fully symmetric in output form or generation regime, which limits strictly matched comparisons on some dimensions.

Future work should therefore extend the benchmark to more language pairs and more realistic interactive settings, strengthen the calibration of speech quality and preservation metrics against human judgments, and further examine how a unified multidimensional protocol can support comparison as speech translation systems continue to evolve.

To support reproducible and extensible research, OpenSTBench will be released as an open-source Python package with a public repository. Users can organize outputs from their own systems according to the shared input format, run the evaluators to obtain multidimensional reports, and replace or extend individual evaluation components for new datasets, language pairs, and research goals.
\section*{Ethical Considerations}

OpenSTBench is intended for research evaluation of speech translation systems. Because the framework includes speaker preservation and voice-cloning-based evaluation resources, it should not be used for speaker verification, impersonation, surveillance, or other identity-sensitive deployment settings. The benchmark is designed to compare system behavior at the aggregate level rather than to identify individual speakers.

We use publicly available datasets, models, and evaluation tools according to their released access conditions and licenses. The LibriTTS-based paired speaker set is constructed for research evaluation only, and derivative artifacts should be used consistently with the licenses and intended use of the original resources. The released OpenSTBench package will document supported datasets, language directions, input/output formats, evaluation components, and intended research use.

The speech datasets used in this work may contain speaker-identifying voice information. We do not attempt to infer real-world identities, collect additional personal information, or release private metadata beyond what is provided by the original datasets. No new human participants were recruited, no new human-subject data collection was conducted, and therefore participant instructions, recruitment, payment, and ethics-board approval are not applicable. We do not introduce new offensive content beyond the original public datasets, and the benchmark does not require collecting or releasing additional user-generated content.

AI assistants were used to support language polishing, checklist preparation, and minor editing suggestions. All technical claims, experimental results, and final manuscript content were reviewed and verified by the authors.

% \section*{Acknowledgments}

\bibliography{custom}

\appendix

\section{Additional Results}

\subsection{Full Speaker Similarity Anchor Results}

\begin{table*}[t]
\centering
\scriptsize
\setlength{\tabcolsep}{4pt}
\begin{tabular}{lcccccccc}
\toprule
& \multicolumn{4}{c}{EN$\to$ZH} & \multicolumn{4}{c}{ZH$\to$EN} \\
\cmidrule(lr){2-5} \cmidrule(lr){6-9}
& \multicolumn{2}{c}{Source anchor} & \multicolumn{2}{c}{Target anchor} & \multicolumn{2}{c}{Source anchor} & \multicolumn{2}{c}{Target anchor} \\
\cmidrule(lr){2-3} \cmidrule(lr){4-5} \cmidrule(lr){6-7} \cmidrule(lr){8-9}
Model
& Resemblyzer$\uparrow$
& WavLM$\uparrow$
& Resemblyzer$\uparrow$
& WavLM$\uparrow$
& Resemblyzer$\uparrow$
& WavLM$\uparrow$
& Resemblyzer$\uparrow$
& WavLM$\uparrow$ \\
\midrule
\multicolumn{9}{l}{\textit{Streaming Models}} \\
Qwen3-LiveTranslate
& 0.5492 & 0.0248 & 0.5967 & 0.2255
& 0.5771 & 0.0833 & 0.5670 & 0.0288 \\
Doubao AST 2.0
& 0.8504 & 0.4516 & 0.8261 & 0.6849
& 0.8461 & 0.5136 & 0.8397 & 0.5392 \\
GPT Realtime Translate
& 0.5498 & 0.0595 & 0.5767 & 0.2275
&  0.6393 & 0.1298 & 0.6389 & 0.1541 \\
\midrule
\multicolumn{9}{l}{\textit{Offline Models}} \\
SeamlessM4T-v2
& 0.5729 & 0.0195 & 0.6086 & 0.2669
& 0.5737 & 0.0881 & 0.5976 & 0.0921 \\
UniSS
& 0.8746 & 0.3575 & 0.8468 & 0.6291
& 0.8776 & 0.4660 & 0.8459 & 0.4619 \\
\bottomrule
\end{tabular}

\caption{Full source-anchor and target-anchor speaker similarity results on the LibriTTS-based paired speaker set. Results are reported separately by direction and model for both Resemblyzer and WavLM. Source anchor compares system output with source-side speech, whereas target anchor compares system output with same-speaker target-language reference speech. Target-anchor comparisons may involve synthesized or natural reference speech depending on the translation direction. Higher is better for all metrics.}
\label{tab:speaker_similarity_anchor_full}
\end{table*}

Table~\ref{tab:speaker_similarity_anchor_full} reports the full speaker-similarity results for the anchor analysis discussed in Section~\ref{sec:speaker_anchor} . The purpose of this analysis is to examine how speaker-preservation scores change under different comparison anchors. In cross-lingual S2ST, comparing generated target speech with source-language speech introduces a language mismatch into the speaker-similarity evaluator. Using target-language reference speech enables same-language comparison, but the comparison condition may still differ depending on whether the anchor speech is natural or synthesized.

We therefore report both source-anchor and target-anchor scores for each system and translation direction. The results show that anchor choice substantially affects the measured similarity, especially for WavLM. The differences reflect not only cross-lingual mismatch, but also natural-versus-synthesized anchor mismatch. For example, in EN$\rightarrow$ZH, target-anchor comparison uses synthesized Chinese reference speech, whereas in ZH$\rightarrow$EN it uses original LibriTTS English speech as the target-language anchor. This helps explain why some ZH$\rightarrow$EN target-anchor WavLM scores remain lower even when the system is designed to preserve speaker characteristics.

These results support our use of same-language target-side anchors for the main speaker-preservation evaluation, while also showing that automatic speaker-similarity scores should be interpreted with care in cross-lingual S2ST.

\section{Metric Definitions}
\label{app:metric-details}

\subsection{Translation Quality}

We report sacreBLEU, chrF++, COMET, and BLEURT for translation quality. Higher values indicate better translation quality.

\subsection{Speech Quality and Preservation}

We report UTMOS for speech naturalness and CER/WER for ASR-based text--speech consistency. Lower CER/WER indicates better consistency.

For speaker preservation, we report Resemblyzer and WavLM speaker similarity. For emotion preservation, we report Emotion2Vec cosine similarity (E2V) and emotion classification accuracy. Higher values indicate stronger preservation.

For paralinguistic fidelity, we report Event Content F1 and Event Timing F1. Event Content F1 evaluates event label and count preservation, while Event Timing F1 additionally considers relative event timing.

\subsection{Time-related Behavior}

For temporal consistency, we report Speech Length Compliant (SLC) score at tolerance thresholds 0.2 and 0.4. Higher SLC indicates better duration consistency.

For streaming latency, we report Start Offset, Average Token Delay (ATD), and Custom ATD. Lower values indicate better responsiveness. For offline local models, we report Real-Time Factor (RTF), where lower values indicate higher efficiency.

\section{Radar Plot Normalization Ranges}
\label{app:radar_ranges}

\begin{table}[t]
\centering
\begin{tabular}{lcc}
\hline
Metric & Direction & Range \\
\hline
BLEU & Higher & [0, 50] \\
chrF++ & Higher & [0, 60] \\
COMET & Higher & [0.5, 0.9] \\
BLEURT & Higher & [0.4, 0.75] \\
UTMOS & Higher & [1, 5] \\
CER/WER & Lower & [0, 0.3] \\
Resemblyzer & Higher & [0, 1] \\
WavLM & Higher & [0, 1] \\
E2V & Higher & [0, 1] \\
Emotion Acc. & Higher & [0, 1] \\
Event Content F1 & Higher & [0, 1] \\
Event Timing F1 & Higher & [0, 1] \\
SLC 0.2 & Higher & [0, 1] \\
SLC 0.4 & Higher & [0, 1] \\
Start Offset & Lower & [0, 8000] ms \\
ATD & Lower & [0, 8000] ms \\
Custom ATD & Lower & [0, 8000] ms \\
RTF & Lower & [0, 2] \\
\hline
\end{tabular}
\caption{Fixed normalization ranges used for radar plots. Direction indicates whether higher or lower raw values are better. Ranges are used only for visualization; main result tables report original metric values.}
\label{tab:radar_ranges}
\end{table}

Table~\ref{tab:radar_ranges} lists the fixed ranges used to normalize metrics in Figure~\ref{fig:radar}. We use fixed ranges rather than per-system min--max normalization so that radar scores are comparable across systems and are not determined by the best and worst systems in the evaluated set. Naturally bounded metrics use their native ranges, while UTMOS follows its MOS-style range. Latency and RTF ranges are clipped for visualization to prevent extreme delays or slow inference from dominating the plot. Translation metrics and CER/WER use broad fixed ranges that cover the empirical values in our experiments. These ranges are used only for visualization; all quantitative comparisons in Tables~\ref{tab:translation_quality}, \ref{tab:speech_attributes}, and \ref{tab:latency_temporal} use the original metric values.

\section{Evaluation Settings}
\label{app:evaluation_settings}

Most evaluation components are used with their released or default inference settings, without additional hyperparameter tuning. We specify model or implementation choices in Section~\ref{sec:openstbench}, including \texttt{Unbabel/wmt22-comet-da} for COMET, the PyTorch implementation of BLEURT, Whisper for ASR transcription, UTMOS for speech quality prediction, WavLM and Resemblyzer for speaker similarity, emotion2vec for emotion similarity, CLAP for acoustic-event detection, and a SimulEval-style interface for latency evaluation.

For settings that require explicit thresholds, SLC is reported with \(\tau=0.2\) and \(\tau=0.4\). CER is used for Chinese, Japanese, and Korean targets, and WER is used otherwise. Latency is computed with each system's native input granularity as described in Section~\ref{sec:evaluation_protocol}. Radar-plot normalization ranges are listed in Appendix~\ref{app:radar_ranges}.

\end{document}